\documentclass{article}
\usepackage{natbib}
\usepackage[accepted]{vietnam}
\usepackage{enumitem}
\usepackage{amssymb}
\usepackage{graphicx}
\usepackage{wrapfig}

\begin{document}
\twocolumn[
\title{Star Formation in the Galactic Center Environment}
\author{Jens Kauffmann}{jens.kauffmann@gmail.com}
\address{Max--Planck--Institut f\"ur Radioastronomie, Bonn, Germany}

\vskip 0.5cm 
]

\begin{abstract}
A brief overview of recent advances in the study of star formation in the Galactic Center (GC) environment is presented. Particular attention is paid to new insights concerning the suppression of star formation in GC molecular clouds. Another focus is the question whether the GC can be used as a template for the understanding of starburst galaxies in the nearby and distant universe: this must be done with care. Some of the particular conditions in the center of the Milky Way do not necessarily play a role in starburst galaxies.
\end{abstract}

\section{The Central Molecular Zone (CMZ)}
About 3--10\% of the total molecular gas and star formation (SF) of the Milky Way reside at $|\ell|\le{}3^{\circ}$ (i.e., within galactocentric radii $\le{}430~\rm{}pc$ ). Unlike the rest of the the Milky Way, the region within $\sim{}200~\rm{}pc$ galactocentric radius (i.e., $|\ell|\lesssim{}1\fdg{}5$) is dominated by gas in molecular instead of atomic form. This domain of the Galaxy is therefore also known as the Central Molecular Zone (CMZ; \citealt{morris1996:cmz-review}).

Many of the most recent results concern the internal structure of CMZ clouds as resolved by interferometers. The global aspects described in the reviews by \citet{gusten1989:review} and \citet{morris1996:cmz-review} thus largely still remain valid today. Much of the current text summarizes material that is discussed in more detail in \citet{kauffmann2016:cmz-review}. Please also refer to that text for many references that must be skipped here due to space restrictions.\medskip

\noindent{}The CMZ is a spectacular star--forming environment. It contains massive and compact young stellar groups like the Arches and Quintuplet clusters that together contain $\gtrsim{}200$ O--type stars and have few --- but still some --- counterparts in the disk of the Milky Way. The CMZ also harbors the Sgr~B2 molecular cloud that alone hosts 59 compact H\textsc{ii}~regions.

Still, relatively few and poor overall constraints on the young stars inhabiting the CMZ exist. It is, e.g., difficult to sense star formation in clouds that are sometimes optically thick even at $\sim{}100~\rm{}\mu{}m$ wavelength. The general CMZ environment resides behind a foreground extinction of $A_{K_{\rm{}s}}\approx{}2~\rm{}mag$ (i.e., $A_V\approx{}18~\rm{}mag$), and so the even the detection of OB~stars outside clouds is difficult. The best constraints on the total star formation activity of the CMZ might therefore come from indirect methods. For example, the number of ionizing photons produced in the CMZ can be estimated from radio data. This constrains the number of high--mass stars emitting such photons, which itself can be related to the star formation rate via further assumptions. Early work (e.g., \citealt{gusten1989:review}: 0.3 to $0.6\,M_{\odot}\,\rm{}yr^{-1}$ at $|\ell|\le{}3\fdg{}5$) is broadly consistent with current estimates (e.g., \citealt{longmore2012:sfr-cmz}: $\le{}0.06\,M_{\odot}\,\rm{}yr^{-1}$ at $|\ell|\le{}1\arcdeg$).\medskip

\noindent{}The distribution of molecular gas in the Galactic Center at $|\ell{}|\le{}5\arcdeg$ is highly asymmetric. About 75\% of the gas seen in $\rm{}^{13}CO$ and CS resides at $\ell{}>0\arcdeg$ and radial velocities $>0~\rm{}km\,s^{-1}$. Interestingly, this distribution is opposite to potential asymmetries in star formation that are possibly seen in \emph{Spitzer} data. \citet{molinari2011:cmz-ring} argue that the gas within $|\ell|\lesssim{}1\arcdeg$ from the Galactic Center forms a connected system of unusually dense and massive molecular clouds on an orbit of $\approx{}100~\rm{}pc$ radius. \citet{henshaw2016:cmz-kinematics} recently demonstrated that the kinematics of these clouds are well explained by an \emph{open} orbit of about 4~Myr period proposed by \citet{kruijssen2014:orbit}: these clouds appear to form a \emph{stream} of clouds that does not close at its ends.

CMZ molecular clouds have unusually high mean $\rm{}H_2$ densities $\sim{}10^4~\rm{}cm^{-3}$ and column densities $\sim{}10^{23}~\rm{}cm^{-2}$. Several clouds concentrate masses $\ge{}10^5\,M_{\odot}$ into just a few parsec radius. The diffuse ionized gas is pervaded by a strong magnetic field $\sim{}10^3~\rm{}\mu{}G$ that also penetrates the CMZ clouds. CMZ molecular clouds have line widths much in excess of Galactic Disk clouds. Many CMZ clouds appear to be subject to violent processes like cloud--cloud collisions at high velocities. This is indicated by widespread emission from SiO and other molecules likely ejected from grain surfaces via shocks, and methanol masers excited in collisions.

Bulk gas temperatures from line ratios are typically in the range 50 to $100~\rm{}K$. It has, however, been speculated that CMZ clouds also contain a cooler and more massive component of gas that is just too faint to dominate the observed line emission. This important topic needs to be revisited in the near future. Gas temperatures $\gtrsim{}50~\rm{}K$ would be mysteriously decoupled from the much lower dust temperatures $\approx{}20~\rm{}K$. This could be explained if gas was heated by agents not affecting the dust, such as cosmic rays. The heating of CMZ gas is an unsolved problem, though. Recent observations suggest that the gas is heated by turbulence \citep{ginsburg2015:cmz-gas-temperatures, immer2016:cmz-temperature}.

It has often been stated that CMZ molecular clouds need to be dense to withstand Galactic Center tides. This argument is, however, based on an outdated understanding of the gravitational potential in the Galactic Center. Current data indicate that tides are actually \emph{compressive} and that the high densities are rather a consequence of the high pressure of the ionized gas in which CMZ clouds are immersed (see Sec.~3 of \citealt{kauffmann2016:cmz-review}).

\section{Suppression of Galactic Center Star Formation}
Given the high gas densities of CMZ clouds, one particularly surprising feature of the region is that \emph{star formation in the dense gas of the CMZ appears to be suppressed}. It was already recognized in the 1980s that, given massive and dense clouds, the CMZ should contain about an order of magnitude more $\rm{}H_2O$ and methanol masers than observed. Our ability to quantify the relation between star formation and dense gas has improved massively over the last few years. Recent studies focusing on the Solar Neighborhood provide a framework against which Galactic Center clouds can be compared. It has now been established that star formation is suppressed on the scale of the entire CMZ (e.g., \citealt{longmore2012:sfr-cmz}) as well as on the scale of individual clouds (e.g., \citealt{kauffmann2013:g0.253, kauffmann2016:gcms_i, kauffmann2016:gcms_ii}). We need to develop a detailed understanding of the conditions in the CMZ in order to understand this suppression of star formation.

Our understanding of the internal structure of CMZ molecular clouds has increased massively over the last few years. The cloud G0.253+0.016 might serve as an example. This region was first discovered in $\rm{}NH_3$ maps of the CMZ published in 1981. In 1994 it was realized that G0.253+0.016 is very extreme in its star formation properties: the cloud concentrates a mass resembling the one of the Orion~A molecular cloud in just $\sim{}3~\rm{}pc$ radius --- but there is no signifiant star formation in this object. A single faint $\rm{}H_2O$ maser, such as expected in regions of low--mass star formation, is the only signpost indicating that young stars exist in this cloud. Subsequent single--dish work further refined the properties of the cloud and its star formation activity. Research on G0.253+0.016 then stopped for several years, given instrumental limitations. \citet{longmore2011:m025} revived this line of work with a fresh look at the object (now a.k.a.\ the ``Brick'') that is primarily motivated by new data from \emph{Herschel}.

None of the aforementioned studies did, however, resolve the internal structure of CMZ clouds. This is a problem: single--dish data probing spatial scales $\gtrsim{}1~\rm{}pc$ constrain how dense molecular cores capable of star formation aggregate out of the diffuse cloud medium. The observations do, however, not reveal the cores themselves on spatial scales $\lesssim{}0.1~\rm{}pc$ where individual stars form. This means that no constraints on the immediate initial conditions for CMZ star formation can be obtained.

Interferometer observations spatially resolving CMZ clouds constitute one of the major recent advances in research exploring Galactic Center star formation. A first study of G0.253+0.016 with the \emph{Submillimeter Array} (\emph{SMA}) by \citet{kauffmann2013:g0.253} reveals a puzzling trend: the cloud has a very high average density, that e.g.\ exceeds that of the Orion~A cloud by an order of magnitude, but \emph{the cloud is essentially devoid of significant dense cores} with radii $\lesssim{}0.1~\rm{}pc$. This trend manifests in rather faint detections of $\rm{}N_2H^+$ in \emph{SMA} maps and an absence of significant dust continuum emission. This result is not a consequence of a low sensitivity of the \emph{SMA} data: cloud cores resembling Orion~KL but located in the CMZ, for example, would be easily detected in such data. Even more detail in G0.253+0.016 is revealed by the \emph{ALMA} data of \citet{rathborne2014:g0253-pdf}. Their dust emission maps confirm the absence of dense cores resembling Orion~KL, i.e., the relative absence of significant dense cores. Similarly, they also show that probability density functions (PDFs) of column density are devoid of excesses at high column density. This is typical for clouds with little star formation activity and quantifies that the cloud is not efficient in concentrating mass at high density. A variety of interferometer--based studies of CMZ clouds have been published in the meantime (see \citealt{kauffmann2016:cmz-review}).\medskip

\noindent{}The aforementioned work gives us a good idea of the spatially resolved properties in a few CMZ clouds. What is now needed is a comprehensive interferometric survey that covers most or all of the CMZ. The Galactic Center Molecular Cloud Survey (GCMS; \citealt{kauffmann2016:gcms_i, kauffmann2016:gcms_ii}) provides a first \emph{SMA}--based overview of the resolved properties in \emph{all} major CMZ clouds except Sgr~B2. This sample, e.g., shows that many CMZ clouds have unusually flat density profiles resembling $\varrho{}\propto{}r^{-1.3}$, much more shallow than a relation $\varrho{}\propto{}r^{-2}$ that would resemble the profiles expected in regions with ongoing star formation. The GCMS data also give information on the kinematics of the dense gas in the clouds. As mentioned before, CMZ molecular clouds as a whole are unusually turbulent, when compared to clouds of similar size in the disk of the Milky Way. The new data do, however, also show that the velocity dispersion of gas in the cores of CMZ clouds of $\lesssim{}0.1~\rm{}pc$ radius are \emph{similar} to what is seen elsewhere in the Milky Way on such small spatial scales. CMZ clouds appear to follow an unusually steep linewidth--size relation.

CMZ star formation is at least in part suppressed because CMZ clouds are inefficient in producing high--mass dense cores of size $\lesssim{}0.1~\rm{}pc$ that could efficiently produce stars. CMZ clouds also exhibit unusual kinematics that need to be studied further.

\section{The CMZ as a Template for Starburst Galaxies}
It is often said that the CMZ might serve as a template for unresolved processes that are active in nearby and more distant starburst galaxies. For example, NGC~253 and the Antennae Galaxies (NGC~4038/39) contain molecular cloud complexes with mean $\rm{}H_2$ column densities $\sim{}10^{23}~\rm{}cm^{-2}$: these regions must be composed of clouds with column densities of the same order, i.e., clouds resembling those in the CMZ. See \citet{kruijssen2013:mw-vs-galaxies} for a related comparison between the Milky Way and other galaxies.

Still, extreme caution is required when using the CMZ as a template for the interpretation of other galaxies. The point is that \emph{several} extreme gas properties potentially influence star formation in this environment. This includes at least the following factors.
\begin{description}[itemsep=0pt,topsep=2pt]
\item[High Gas Temperatures] The high CMZ gas temperatures $T_{\rm{}gas}\ge{}50~\rm{}K$ imply high Bonnor--Ebert masses $m_{\rm{}BE}=20\,M_{\odot}\cdot{}(T_{\rm{}gas}/50~{\rm{}K})^{3/2}\cdot{}(n/10^5~{\rm{}cm^{-3}})^{-1/2}$ for gas at $\rm{}H_2$ particle density $n$. This is by a factor $\approx{}6$ higher than for typical Solar Neighborhood clouds, where $T_{\rm{}gas}\approx{}15~\rm{}K$. The high value of $m_{\rm{}BE}$ might help to suppress star formation.
\item[High Gas Velocity Dispersion] The high gas velocity dispersion in CMZ clouds can provide further support against collapse. This effectively further increases $m_{\rm{}BE}$. Note, however, that the velocity dispersion in CMZ cloud fragments of $\lesssim{}0.1~\rm{}pc$ radius appears to resemble what is found elsewhere in the Milky Way. The gas velocity dispersion might thus not have an effect particular to CMZ star formation.
\item[Strong Magnetic Field] The magnetic field in CMZ clouds is unusually strong. This should again further increase the effective value of $m_{\rm{}BE}$.
\item[High Gas Densities] At the same time the high pressure $\sim{}5\times{}10^6~\rm{}K\,cm^{-3}$ from hot ionized gas confines the clouds. Given gas temperatures $\sim{}100~\rm{}K$, the gas in clouds must reside at densities $\gtrsim{}10^4~\rm{}cm^{-3}$. These high densities might improve conditions for star formation.
\item[Frequent Perturbations \& Cloud--Cloud Collisions] CMZ clouds orbit the GC environment about every 4~Myr. The clouds are thus subject to frequent and regular perturbations. Given that clouds are tightly packed on this orbit, it also seems plausible that cloud--cloud collisions at high relative velocity occur often. This is indeed indicated by strong SiO emission in the CMZ. Future theoretical research should examine whether these effects should promote or suppress star formation.
\end{description}
The interplay of these and potentially other factors needs to be studied when using the CMZ as a template for parts of other galaxies. Some starburst galaxies might, e.g., contain gas at densities and temperatures resembling values found in the CMZ. But those extragalactic clouds might potentially reside far away from the galaxy's center. In that case there are no frequent and strong perturbations due to orbital motion. Then we could use the CMZ as a laboratory to study the physics of star formation at densities similar to those in starburst galaxies. But this research will only be useful if we remove the impact of short orbit periods from the observed trends. We thus must strive to understand the implications of all the particular physical effects acting in the CMZ.


\end{document}